\documentstyle[12pt,a4,fleqn,epsfig]{article}
\normalbaselineskip=\baselineskip

\begin{document}
%
%
\begin{titlepage}
\vspace*{-1cm}
\phantom{bla}
\hfill{DFPD 96/TH 30}
\\
\phantom{bla}
\hfill{hep-ph/}
\vskip 1cm
\begin{center}
{\huge \bf 
Fermion Virtual Effects in \\
\vskip 0.2cm
$e^+e^- \rightarrow W^+W^-$ Cross Section
\footnote{Work partially supported by the European Union
under contract No.~CHRX--CT92--0004, and by MURST.}
}
\end{center}
\vskip 2.0cm

\begin{center}
\centerline{
$\mbox{A. CULATTI}^\dagger$\footnote{
\noindent
Present address Dep. de Fisica Teorica y del Cosmos, Universidad de Granada, 
Spain.}, 
$\mbox{G. DEGRASSI}^\dagger$, $\mbox{F. FERUGLIO}^\dagger$, 
$\mbox{A. MASIERO}^\ddagger$,}
\centerline{$\mbox{S. RIGOLIN}^\dagger$, 
$\mbox{L. SILVESTRINI}^\ast$, $\mbox{A. VICINI}^\dagger$}
\vskip 0.5truecm
\centerline{$\mbox{ }^\dagger$  {\it Dipartimento di Fisica - Universit\`a di 
Padova and INFN Padova, Italy}}
\centerline{$\mbox{ }^\ddagger$ {\it Dipartimento di Fisica - Universit\`a di 
Perugia and INFN Perugia, Italy}}
\centerline{$\mbox{ }^\ast$ {\it Dipartimento di Fisica - Universit\`a di 
Roma II "Tor Vergata" and INFN Roma II, Italy}}
\vskip .2cm
\end{center}

\vskip 1.5cm

\begin{abstract}
\noindent
We analyse the contribution of new heavy virtual fermions to the 
$e^+e^- \rightarrow W^+W^-$ cross section. We find that there exists a relevant 
interplay between trilinear and bilinear oblique corrections. The result strongly 
depends on the chiral or vector--like nature of the new fermions. 
As for the chiral case we consider sequential fermions: one obtains 
substantial deviation from the Standard model prediction, making the effect 
possibly detectable at $\sqrt{s}=500$ or $1000$ GeV linear colliders. 
As an example for the vector--like case we take a SUSY extension with heavy 
charginos and neutralinos: due to cancellation, the final effect turns 
out to be negligible. 
\end{abstract}

\vfill{
\noindent
Contribution at {\it ``Physics with $e^+ e^-$ Colliders''} (The European Working Groups, 
4 Feb 1995 - 1 Sep 1995), Hamburg, Germany, 30 Aug - 1 Sep 1995 and at 
{\it ``3rd Workshop on Physics and Experiments with $e^+ e^-$ Linear Colliders''}, 
Iwate, Japan, 8-12 Sep 1995. 
}

\end{titlepage}
\setcounter{footnote}{0}



\def    \bc            {\begin{center}}
\def    \ec            {\end{center}}
\def    \DD            {{\cal D}}
\def    \LL            {{\cal L}}
\def    \dmu           {\partial_\mu}
\def    \dnu           {\partial_\nu}
\def    \dd            {\displaystyle}
\def    \be            {\begin{equation}}
\def    \ee            {\end{equation}}
\def    \bea           {\begin{eqnarray}}
\def    \eea           {\end{eqnarray}}
\def    \nn            {\nonumber}
\def    \Asm           {\mbox{$ A_{\lambda{\bar\lambda}}^{SM} $}}
\def    \Ag            {\mbox{$ A_{\lambda{\bar\lambda}}^\gamma $}}
\def    \Az            {\mbox{$ A_{\lambda{\bar\lambda}}^Z $}}
\def    \dAg           {\mbox{$ \delta A_{\lambda{\bar\lambda}}^\gamma $}}
\def    \dAz           {\mbox{$ \delta A_{\lambda{\bar\lambda}}^Z $}}
\def    \DAg           {\mbox{$ \Delta A_{\lambda{\bar\lambda}}^\gamma $}}
\def    \DAz           {\mbox{$ \Delta A_{\lambda{\bar\lambda}}^Z $}}
\def    \M2            {\mbox{$ \tilde{\cal M} $}}
\def    \Mg            {\mbox{$ {\tilde{\cal M}}^\gamma $}}
\def    \Mz            {\mbox{$ {\tilde{\cal M}}^Z $}}
\def    \Mnu           {\mbox{$ {\tilde{\cal M}}^\nu $}}
\def    \thetab        {\bar\theta}
\def    \raw           {\rightarrow}
\def    \law           {\leftarrow}
\def    \lraw          {\leftrightarrow}
\def    \lum           {fb^{-1}}
\def    \sb            {\mbox{$ {\bar{s}} $}}
\def    \cb            {\mbox{$ {\bar{c}} $}}


We present here a study of virtual effects from
new physics in the cross--section for $e^+e^-\to W^+W^-$.
This analysis has been motivated by the following questions. What room is left
 for deviations with respect 
to the SM predictions, once the constraints coming from the 
LEP1 results have been properly accounted for?
Are these deviations entirely due to the appearance of anomalous
trilinear vector--boson couplings, as generally assumed? 
Is there any role played by the so--called oblique corrections?

To deal with these questions we have focused on two simple SM
extensions:

\noindent
$\bullet$ Model 1: {\bf the SM plus heavy chiral fermions}. In particular we 
consider an extra doublet of heavy quarks, 
exact replica of the SM counterparts, as far as their electroweak and strong
quantum numbers are concerned;

\noindent
$\bullet$ Model 2: {\bf the SM plus heavy vector--like fermions}. As an example 
we take the Minimal Supersymmetric Standard Model (MSSM) with  heavy 
(above the production threshold) electroweak gauginos and higgsinos, 
and very heavy (decoupled) squarks, sleptons and additional higgses.

In both models the one--loop corrections to the process in question are 
concentrated in vector--bosons self--energies and three--point functions, 
which makes it possible to analyse the interplay between the two 
contributions \cite{holdom}.
Box corrections are obviously absent in the first model, while they
are negligibly small in the second case, provided that the scalar 
masses are sufficiently large. Then the only relevant contributions
remain those of gauginos and higgsinos to vertices and self--energies.

The constraints coming from LEP1 results are quite different in the 
two models.
A mass splitting between the two new quarks, whose left--handed components
are partners of a common $SU(2)_L$ doublet, results in a potentially
large contribution to the $\epsilon_1$ parameter. To avoid disagreement
with the experimental result \cite{fit}, we will consider the case 
of degenerate quarks (a detailed discussion concerning this point is
provided in \cite{fmrs}). 
A more stringent limitation comes from the $\epsilon_3$ variable,
which depends mildly on the quark masses. Each new degenerate
quark doublet contributes to $\epsilon_3$ with the positive amount
\be
\delta\epsilon_3=\frac{G_F m_W^2}{4 \pi^2 \sqrt{2}}\simeq 1.3\cdot 10^{-3}.
\label{v1}
\ee
For $m_t=175$ GeV and $m_H=100$ GeV, a single new coloured doublet (with
the possible addition of a lepton doublet) 
saturates the present $2 \sigma$ upper bound on $\epsilon_3$,
and additional chiral doublets beyond the multiplet considered 
here are experimentally ruled out.

On the contrary, the version of the MSSM here analysed easily respects
the constraints from LEP1.
Indeed contributions to the $\epsilon$ variables come in inverse power
of the gaugino and higgsino masses and they turn out to be 
numerically quite small, within the experimental bound.
For practical purposes we have considered the case of degenerate
electroweak gauginos of mass $M$ and degenerate higgsinos of
mass $\mu$. This is the case when the mixing
terms among gauginos and higgsinos are much smaller compared to
the SUSY breaking gaugino mass term $M$ and the supersymmetric
$\mu$ contribution coming from the $H_1 H_2$ term in the superpotential.

After the inclusion of the one--loop corrections due to the new particles
and of the appropriate counterterms, the reduced amplitude for the 
process at hand reads, following the conventions of ref.
\cite{hpz}:

\noindent
$\bullet \Delta\lambda=\pm2$
\be
{\tilde{\cal M}}=-\frac{\sqrt{2}}{\sin^2\thetab}
\delta_{\Delta\sigma,-1} \left[1-\frac{\sin^2\thetab}{\cos 2\thetab} 
\Delta r_W-e_6\right]
\frac{1}{1+\beta^2-2 \beta \cos\Theta}
\label{v2}
\ee

\noindent
$\bullet \Delta\lambda=\le 1$
\bea
\Mg &=&-\beta \delta_{|\Delta\sigma|,1} \left[1+\Delta\alpha(s)\right]
\left[\Ag +\dAg (s)\right]\nn\\
\Mz &=&\beta \frac{s}{s-m_Z^2} 
\left[\delta_{|\Delta\sigma|,1}-\frac{\delta_{\Delta\sigma,-1}}{2\sin^2\thetab
(1+\Delta k(s))}\right] \cdot \nn \\
   & &~~~~~~~~~~~~~~~~~~~~~~~~\left[1+\Delta\rho(s)+
     \frac{\cos2\thetab}{\cos^2\thetab}\Delta k(s)\right]
     \left[\Az +\dAz (s)\right]\nn\\
\Mnu &=&\frac{\delta_{\Delta\sigma,-1}}{2\sin^2\thetab~\beta}
        \left[1-\frac{\sin^2\thetab}{\cos 2\thetab} 
        \Delta r_W-e_6\right]\left[B_{\lambda{\bar\lambda}}-
        \frac{1}{1+\beta^2-2 \beta \cos\Theta} C_{\lambda{\bar \lambda}}\right]
\label{v3}
\eea

In eq. (\ref{v3}) $\beta=(1-4 m_W^2/s)^{1/2}$, $\Theta$ is the scattering angle of 
$W^-$ with respect to $e^-$ in the $e^+e^-$ c.m. frame; $\sigma$, $\bar\sigma$, 
$\lambda$, $\bar\lambda$ are the helicities for $e^-$, $e^+$, $W^-$ and $W^+$, 
respectively; $\Delta\sigma=\sigma-{\bar\sigma}$; $\Ag$, $\Az$, 
$B_{\lambda{\bar\lambda}}$ and $C_{\lambda{\bar\lambda}}$ are tree--level SM 
coefficients listed in Table 1; $\Delta\alpha(s)$, $\Delta k(s)$, $\Delta\rho(s)$,  
$\Delta r_W$ and $e_6$ are finite self--energy corrections \cite{bfc,eps}, given
in the Appendix where also the effective weak angle $\thetab$ is defined.
\begin{table}[t]
\centering
\begin{tabular}{||c|c|c|c||} \hline
& & & \\
$\lambda\bar\lambda$ & $\Ag = \Az$ & $B_{\lambda \bar\lambda}$ & 
                               $C_{\lambda \bar\lambda}$ \\
& & & \\ \hline
$++ \, ,\,-- $ & $ 1 $ & 1 & $1/ \gamma^2$ \\
$+0 \, ,\,0-$ & $ 2 \gamma $ & $2\gamma$ & $2(1+\beta)/\gamma$ \\
$0+ \, ,\,-0$ & $2 \gamma $ & $2\gamma$ & $2(1-\beta)/\gamma$ \\
$00$ & $2 \gamma^2 + 1$ & $2\gamma^2$ & ${2/ \gamma^2}$ \\ \hline
\end{tabular}
\caption{ 
\footnotesize 
Standard Model coefficients expressed in terms of $\gamma^2=s/4 m_W^2$.}
\end{table}
Finally, $\dAg$ and $\dAz$ represent the corrections to the trilinear
gauge boson vertices. For the models considered, they can be expressed 
in terms of the $CP$--invariant form factors $\delta f_i^{\gamma,Z}~~~
(i=1,2,3,5)$ according to the relations:
\bea
\delta A_{++}^V&=&\delta A_{--}^V=\delta f_1^V\nn\\
\delta A_{+0}^V&=&\delta A_{0-}^V=\gamma(\delta f_3^V + \beta f_5^V)\nn\\
\delta A_{-0}^V&=&\delta A_{0+}^V=\gamma(\delta f_3^V - \beta f_5^V)\nn\\
\delta A_{00}^V&=&\gamma^2\left[-(1+\beta^2) \delta f_1^V + 4 \gamma^2 \beta^2 
\delta f_2^V + 2 \delta f_3^V\right]
\label{v4}
\eea                             
Here $\delta f_i^V$ includes both the contribution coming from the 1PI one--loop 
correction to the vertex $VWW$ and the wave--function renormalization of the external 
$W$ legs, taken on the mass--shell. This makes the terms $\delta f_i^V$ finite. 
Other choices in the renormalization conditions are equivalent (see for example 
\cite{peskin}).

Some comments are in order. The tree--level SM amplitudes are recovered
from the above formulae by taking $\Delta\alpha(s)=\Delta k(s)=\Delta\rho(s)=
\Delta r_W=e_6=\dAg=\dAz=0$.
In the high--energy limit, the individual SM amplitudes
from photon, $Z$ and $\nu$ exchange are proportional to $\gamma^2$
when both the $W$'s are longitudinally polarized ($LL$) and proportional
to $\gamma$ when one $W$ is longitudinal and the other is transverse
($TL$). The cancellation of the $\gamma^2$ and $\gamma$ terms
in the overall amplitude is guaranteed by the tree--level, asymptotic relation
$\Ag=\Az=B_{\lambda{\bar\lambda}}$.
When one--loop contributions are included, one has new terms 
proportional to $\gamma^2$ and $\gamma$ (see $\dAg$ and $\dAz$ in eq.
(\ref{v4})) and the cancellation of those terms in the high--energy limit entails
relations among oblique and vertex corrections. We have explicitly
checked that in all cases considered, this cancellation does take
place. We stress that omitting, for instance, the gauge boson self--energies
such cancellation does not occur any longer and the resulting amplitudes
violate the requirement of perturbative unitarity. 

On the other hand, one of the possibilities one can think of
to have appreciable deviations in the cross--section is to delay 
the behaviour required by unitarity \cite{peskin}.
This may happen if in the energy window $m_W<<\sqrt{s}\le 2 M$
($M$ denoting the mass of the new particles)
the above cancellation is less efficient and terms proportional
to positive power of $\gamma$ survive in the total amplitude.
If $\gamma$ is sufficiently large, then a sizeable deviation from the SM 
prediction is not unconceivable.

We now come to the quantitative discussion.

\noindent
$\bullet$ Model 1: {\bf Chiral fermions}. 

In fig. 1 we plot N, the number of  $W^+W^-$ events per bin, versus $\Theta$ 
at $\sqrt{s}=500$ GeV in the channel LL, taking $M=300$ GeV and assuming a 
luminosity of $20~{\rm fb}^{-1}$. 
%
%
\begin{figure}[t]
\label{sample1}
\vspace{0.1cm}
\centerline{\epsfig{figure=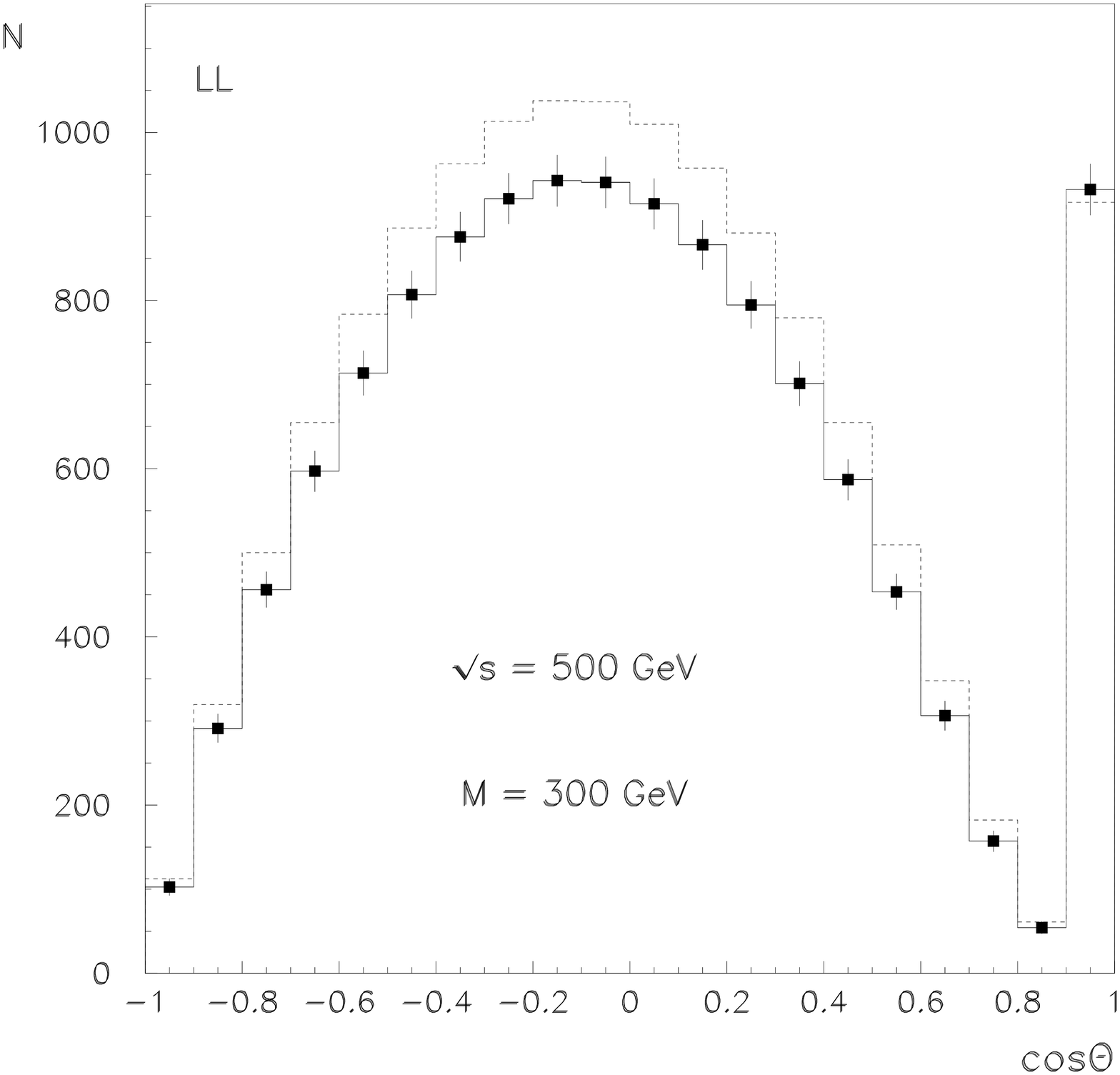,height=10cm,width=11cm,angle=0}}
\caption{\footnotesize 
Number of  $W^+W^-$ events per bin versus $\cos\Theta$ at 
$\sqrt{s}= 500$ GeV in the LL channel, taking $M=300$ GeV and assuming a 
luminosity of $20~{\rm fb}^{-1}$.}
\end{figure}
The error bars refer to the statistical error, the full line  denotes the SM
expectation at tree level\footnote{SM corrections shift both curves in 
fig. 1 by the same amount. Hence, although they change the number N, 
they do not affect our result concerning the size of the deviation due 
to the presence of new physics.}, while the dashed line gives the prediction 
for model 1.
We notice a clear indication of a significant signal, fully
consistent with present experimental bounds. The departure from the SM 
prediction \cite{arg} becomes even more conspicuous for larger $\sqrt{s}$. 
In fig.2 we report the result for  $\sqrt{s}=1000$ GeV, $M=600$ GeV and a 
luminosity of $100~{\rm fb}^{-1}$. 

%
%
\begin{figure}[t]
\label{sample2}
\vspace{0.1cm}
\centerline{
\epsfig{figure=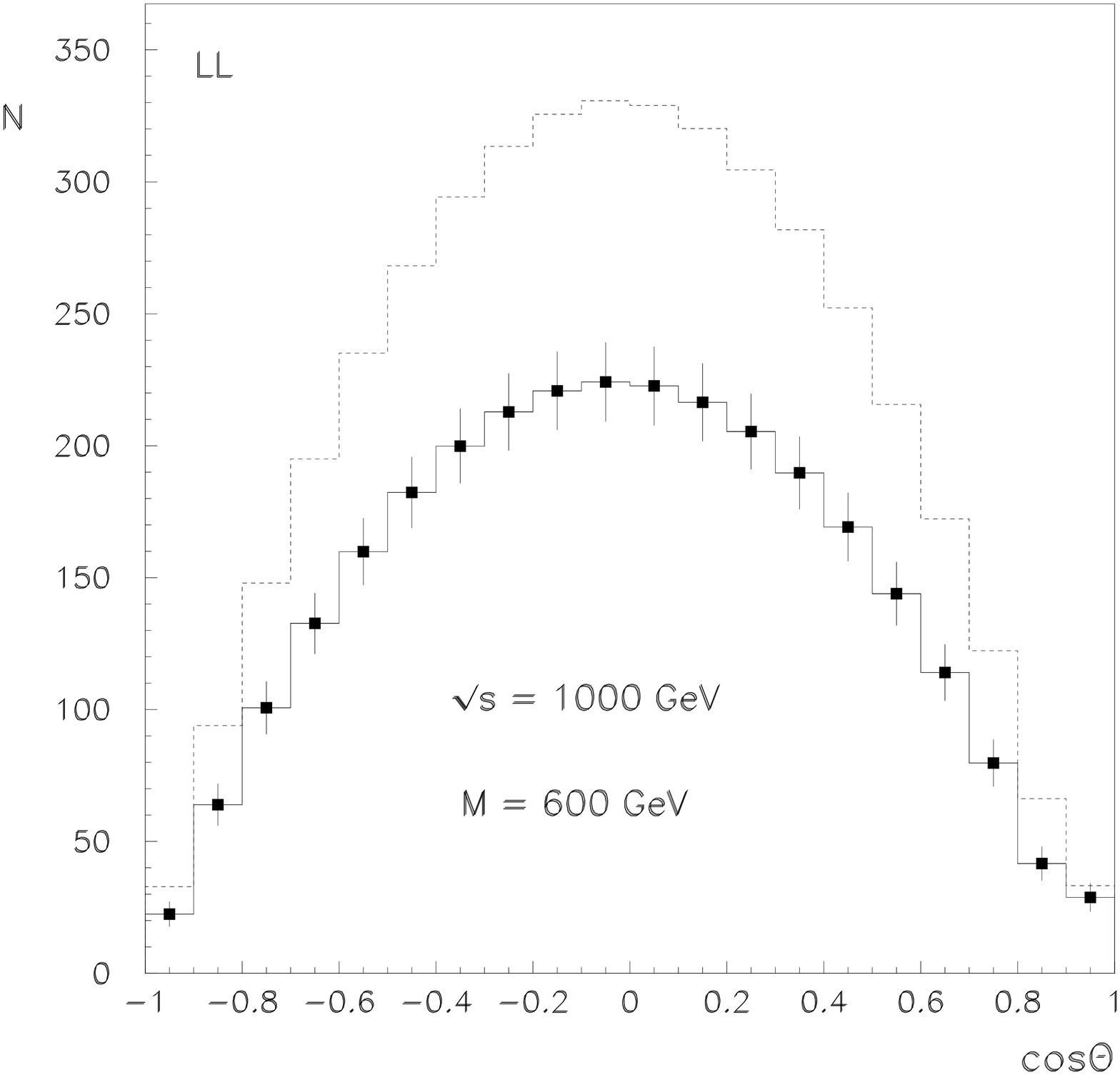,height=10cm,width=11cm,angle=0}
}
\caption{\footnotesize 
Number of  $W^+W^-$ events per bin versus $\cos\Theta$ at 
$\sqrt{s}=1000$ GeV in the LL channel, taking $M=600$ GeV and assuming a 
luminosity of $100~{\rm fb}^{-1}$.}
\end{figure}

For understanding better this point it is useful introduce the following quantity 
$\Delta R$
\be
\Delta R=\frac{\dd\left(\frac{d\sigma}{d\cos\Theta}\right)-
\dd\left(\frac{d\sigma}{d\cos\Theta}\right)_{SM}}
{\dd\left(\frac{d\sigma}{d\cos\Theta}\right)_{SM}}.
\label{v9}
\ee
that represents the deviation between the SM and new physics differential 
cross--section normalized to the SM. 
For $\cos\Theta$ not close to 1, and in the limit  $m_W<<M,\sqrt{s}$ we have derived 
the following analytical expression for $\Delta R_{LL}$:
\be
\Delta R_{LL}=|1+2 K|^2-1
\label{v10}
\ee
\be
K=\frac{g^2 N_c}{16 \pi^2} \frac{4 M^2}{s}\cdot \gamma^2 \cdot
F\left(\frac{M^2}{s}\right), 
\label{v11}
\ee
\be
F(x)=1-\sqrt{4x-1}~\arctan\frac{1}{\sqrt{4x-1}},
\label{v12}
\ee

For $s>>M^2$, $F$ grows only logarithmically and unitarity is respected.
When $M^2>>s$, $F\simeq s/12 M^2$ and the decoupling property is
violated, as one expects in the case of heavy chiral fermions.
In the range of energies we are interested in, $m_W<<\sqrt{s}\le 2M$,
$K$ is of order $G_F M^2$, which explains the magnitude of the 
effect exhibited in fig. 1 and 2. A similar behaviour is also exhibited by
the $TL$ channel.
Indeed, the absolute deviation in the $TL$ channel is of $O(\gamma)$, 
but an additional $\gamma$ factor comes from the SM cross section in the 
denominator of $\Delta R$. In fig. 3 we depict the quantity $\Delta R$ 
as a function of $\cos\Theta$ for LL channel and for the  unpolarized cross 
section at $\sqrt{s}=500$ GeV. 
Notice that the effect which is so large in the $LL$ channel for all the angles
essentially disappears in the unpolarized cross section for 
$\cos\Theta \simeq 1$ where bigger is the number of events expected. 
This makes it clear that for the observation of the effect under discussion 
good identification of the $W$ polarization and large luminosity are 
essential.


%
%
\begin{figure}
\centerline{
\begin{tabular}{cc}
\epsfig{figure=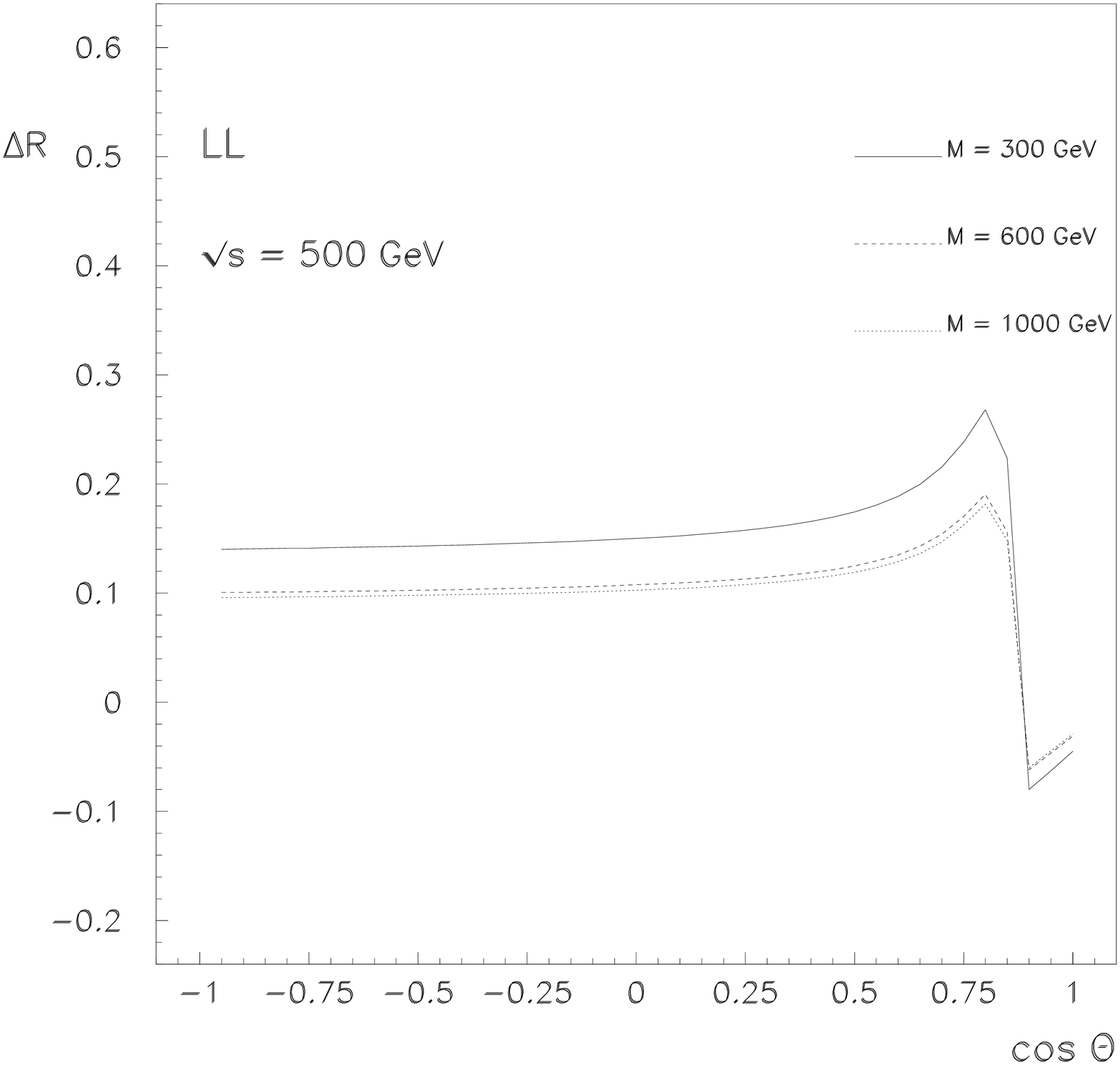,height=8cm,angle=0} & 
\epsfig{figure=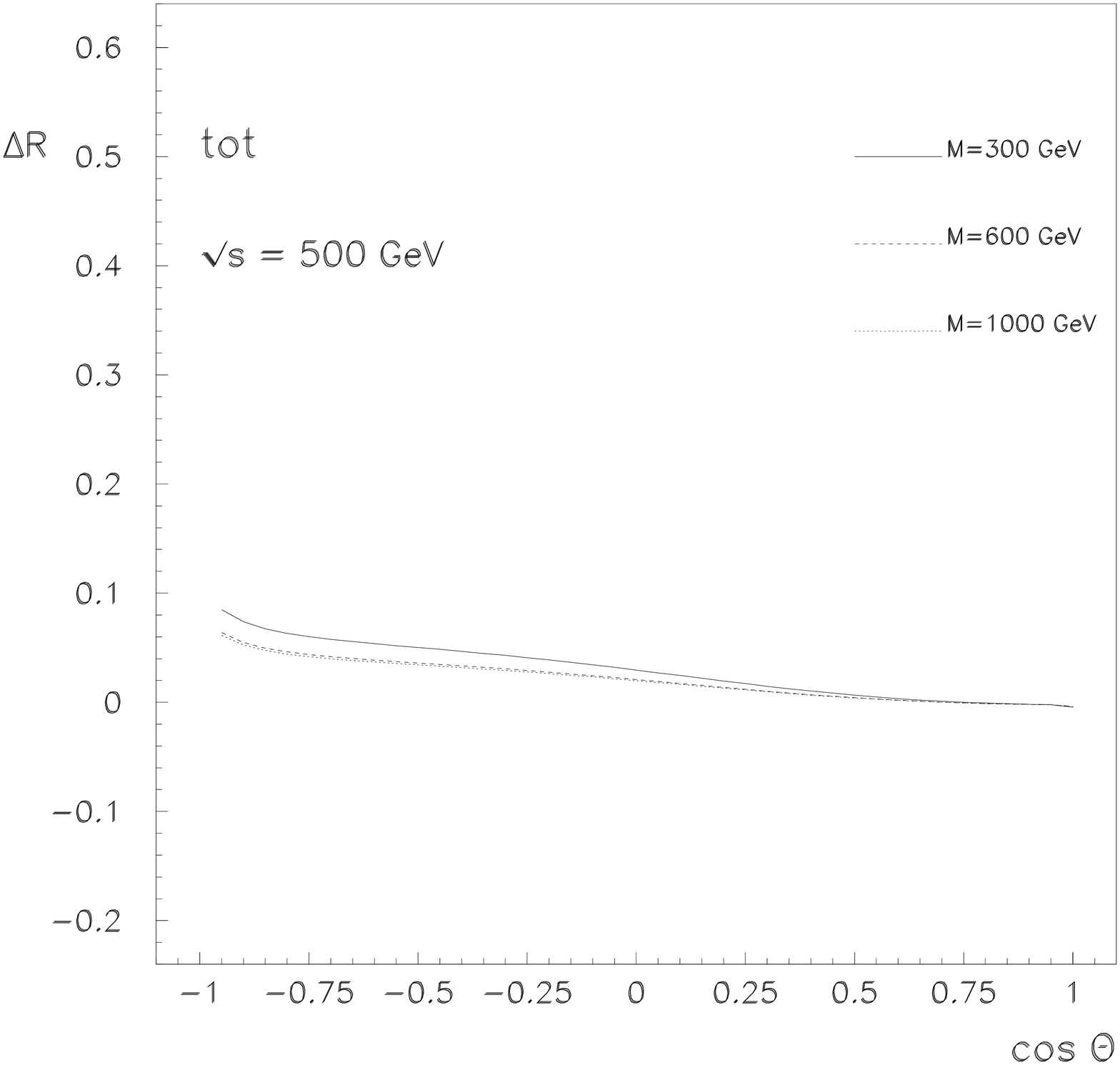,height=8cm,angle=0} \\
\end{tabular}
}
\caption{\footnotesize
Relative deviation $\Delta R$ versus $\cos\Theta$ in the LL polarization 
channel and for the unpolarized cross section (tot) at $s=(500 \mbox{ GeV})^2$ 
and  $M=300, 600, 1000 $ GeV for model 1.}
\label{sample3}
\end{figure}

\noindent
$\bullet$ Model 2: {\bf Vector--like fermions}.

In this case 
the deviations at $\sqrt{s}=500$ GeV or even higher energies
reach at most a few per mille  making it questionable to have a chance 
at all to observe the effect \cite{ls}. 
The analytic expressions for $\Delta R$ both in the  $LL$
and in the $TL$ channel, in the limit $m_W<<M,\sqrt{s}$, are 
vanishing, showing that in the case of gauginos or higgsinos,
which have vector--like coupling, no unitarity delay takes place.
This is in agreement with the decoupling theorem \cite{app}.

To exemplify the danger of including in the cross--section
only the contribution of trilinear anomalous gauge couplings,
we have computed the $\Delta R_{LL}$ ratio at $\sqrt{s}=500~GeV$, 
for model 2, $M=300$ GeV and $\mu>>M$, by leaving out
the contribution from the vector boson self--energies. Due to
the improvident omission of an essential part of the one--loop
correction, one gets an irrealistic $\Delta R_{LL}\simeq-0.2$, with failure of 
unitarity when higher energies are considered.

\section*{Appendix}

The quantities $\Delta\alpha(s)$, $\Delta k(s)$, $\Delta\rho(s)$,  
$\Delta r_W$ and $e_6$ appearing in eqs. (\ref{v2}) and (\ref{v3})
are defined by the following relations in terms of the unrenormalized
vector-boson vacuum polarizations:
\bea
e_1&=&\frac{\Pi_{ZZ}(0)}{m_Z^2}-\frac{\Pi_{WW}(0)}{m_W^2}\nn\\
e_2&=&\Pi'_{WW}(0)-\cos^2\thetab~
\Pi'_{ZZ}(0)-2\cos\thetab\sin\thetab~
\frac{\Pi_{\gamma Z}(m_Z^2)}{m_Z^2}\nn\\
 & & - \sin^2\theta~\Pi'_{\gamma\gamma}(m_Z^2)\nn\\
e_3(s)&=&\frac{\cos\thetab}{\sin\thetab}\left\{\sin\thetab\cos\thetab\left[
\Pi'_{\gamma\gamma}(m_Z^2)-\Pi'_{ZZ}(0)\right]+
\cos 2\thetab~ \frac{\Pi_{\gamma Z}(s)}{s}\right\}\nn\\
e_4&=&\Pi'_{\gamma\gamma}(0)-\Pi'_{\gamma\gamma}(m_Z^2)\nn\\
e_5(s)&=&\Pi'_{ZZ}(s)-\Pi'_{ZZ}(0)\nn\\
e_6&=&\Pi'_{WW}(m_W^2)-\Pi'_{WW}(0)\\
\Delta\alpha(s)&=&\Pi'_{\gamma\gamma}(0)-\Pi'_{\gamma\gamma}(s)\nn\\
\Delta k(s)&=&-\frac{\cos^2\thetab}{\cos 2\thetab}~ (e_1-e_4)+
              \frac{1}{\cos 2\thetab}~ e_3(s)\nn\\
\Delta\rho(s)&=&e_1-e_5(s)\nn\\
\Delta r_W&=&-\frac{\cos^2\thetab}{\sin^2\thetab}~e_1+
\frac{\cos 2\thetab}{\sin^2\thetab}~e_2 + 2~e_3(m^2_Z) + e_4 \\
\epsilon_1&=&e_1-e_5(m_Z^2)\nn\\
\epsilon_2&=&e_2-\sin^2\thetab~ e_4 -\cos^2\thetab ~e_5(m_Z^2)\nn\\
\epsilon_3&=&e_3(m_Z^2)+\cos^2\thetab~ e_4-\cos^2\thetab~ e_5(m_Z^2)
\eea
with
\be
\Pi'_{VV'}(s)=\frac{\Pi_{VV'}(s)-\Pi_{VV}(m_{VV'}^2)}{(s-m_{VV'}^2)}
~~~~~~~~~{\rm with}~~~(V,V'=\gamma,Z,W)
\ee
and $m_{\gamma \gamma}=m_{\gamma Z}=0$, $m_{ZZ}=m_Z$ and $m_{WW}=m_W$.
\label{v8} 
If $s=m_Z^2$ then $\Delta\alpha(s)$, $\Delta k(s)$, $\Delta\rho(s)$
coincides with the corrections $\Delta\alpha$, $\Delta k$, $\Delta\rho$
which characterize the electroweak observables at the $Z$ resonance.
Finally, the effective weak angle $\thetab$ is defined by:
\be
\sin^2\thetab=\frac{1}{2}-\sqrt{\frac{1}{4}-\frac{\pi\alpha(s)}{\sqrt{2} G_F
m_Z^2}}
\ee
where $\alpha(s)$ is the electromagnetic coupling with all the effects 
coming from SM particles included at the given energy $s$.

%

\end{document}